%% file: ANA-TOPQ-2023-32-PAPER.tex
\pdfoutput=1
\documentclass[PAPER, cernpreprint, texlive=2023, UKenglish, texmf, orcidlogo]{atlasdoc}

\usepackage{todonotes}
\DeclareBoolOption[true]{LANGEDIT}
\LEcontact{Chip Brock brockr@msu.edu}
\newcommand*{\LEinline}[1]%
{\todo[inline, color=yellow!30, size=\footnotesize]{#1}}

\usepackage[subfigure=true]{atlaspackage}
\usepackage{atlasbiblatex}

\usepackage{atlasphysics}
\graphicspath{{logos/}{figures/}}

\usepackage{ANA-TOPQ-2023-32-PAPER-defs}

\AtlasTitle{Observation of $t\bar{t}$ production in the lepton+jets and dilepton channels
in $p$+Pb collisions at $\sqrt{s_\mathrm{NN}}=8.16$~\tev\ with the ATLAS detector}

\AtlasAbstract{%
This paper reports the observation of top-quark pair production
in proton-lead collisions in the ATLAS experiment at the Large Hadron Collider.
The measurement is performed using 165~nb$^{-1}$ of $p$+Pb data
collected at $\sqrt{s_\mathrm{NN}}=8.16$~\ensuremath{\text{TeV}}\xspace in 2016.
Events are categorised in two analysis channels, consisting of either events with exactly
one lepton (electron or muon) and at least four jets, or events with two opposite-charge
leptons and at least two jets. In both channels at least one $b$-tagged jet is also required.
Top-quark pair production is observed with a significance over five standard deviations in each channel.
The top-quark pair production cross-section is measured to be
$\sigma_{t\bar{t}}= 58.1\pm 2.0\;\mathrm{(stat.)\;^{+4.8}_{-4.4} \;\mathrm{(syst.)}}\;\mathrm{nb}$,
with a total uncertainty of 9\%.
In addition, the nuclear modification factor is measured to be
$R_{p\mathrm{A}} = 1.090\pm0.039\;(\mathrm{stat.})\;^{+0.094}_{-0.087}\;(\mathrm{syst.})$.
The measurements are found to be in good agreement with theory predictions involving
nuclear parton distribution functions.
}

\AtlasRefCode{TOPQ-2023-32}

\PreprintIdNumber{CERN-EP-2024-097}

\AtlasJournalRef{\JHEP 11 (2024) 101}
\AtlasDOI{10.1007/JHEP11(2024)101}

\hypersetup{pdftitle={ATLAS document},pdfauthor={The ATLAS Collaboration}}

\begin{document}

\maketitle

\tableofcontents
\section{Introduction}
The heavy-ion~(HI) collisions produced at the TeV-scale energies of the Large Hadron Collider~(LHC)
at CERN opened up the possibility to measure various elementary particle production
for the first time in lead-lead~(Pb+Pb) and proton-lead~(\pPb) systems.
With the first observations of the $W^\pm$~boson~\cite{ATLAS:2014sic,CMS:2012fgk},
the $Z$~boson~\cite{ATLAS:2010xzb,ATLAS:2012qdj,CMS:2011zfr},
bottom-quark jets ($b$-quark jets)~\cite{ATLAS:2022agz,CMS:2013qak}
and the $\tau$-lepton~\cite{ATLAS:2022ryk,CMS:2022arf},
only two Standard Model~(SM) particles remain
to be directly observed in Pb+Pb collisions: the Higgs boson and the top quark.
While the Higgs boson production cross-section is too low to establish its observation
in Pb+Pb collisions at the LHC, the observation of top-quark production is within reach.

In ultra-relativistic Pb+Pb collisions at the LHC, top quarks are expected to provide
a unique tool to measure properties of the strongly interacting quark-gluon plasma~(QGP)~\cite{Apolinario:2017sob}.
These properties may be inferred from modifications of various observables
in Pb+Pb collisions in comparison to reference measurements in the proton--proton~($pp$) system.
Often, modifications of event yields in Pb+Pb collisions relative to reference yields in $pp$ collisions
can be attributed to both initial-state effects (e.g., different parton distribution functions for
heavy nuclei than for free nucleons) and final-state effects due to the creation of the QGP\@.
Therefore, precise knowledge of the initial-state effects is crucial in anticipation of
extracting QGP properties precisely from experimental data.
This knowledge may be gained using measurements done in \pPb\ collisions.

In \pPb\ collisions, top quarks provide novel probes of nuclear
parton distribution functions~(nPDFs)~\cite{dEnterria:2015mgr,Khanpour:2016pph,Ethier:2020way},
especially the gluon nPDF, which is particularly important for perturbative
calculations in Quantum Chromodynamics~(QCD) at the LHC energies~\cite{Paukkunen:2014nqa,Paukkunen:2017bbm,Ethier:2020way}.
The top-quark yields, measured using kinematics of electrons and muons originating from top-quark decays,
provide precise information~\cite{dEnterria:2015mgr}
in the kinematic region of Bjorken-$x\sim 3\cdot 10^{-3}-0.5$ and
$Q^2\sim m_\mathrm{t}^2\sim 3 \cdot 10^4$~$\GeV^2$
($m_\mathrm{t}$ stands for the top-quark mass), which is poorly constrained by other measurements.
In this region, anti-shadowing and EMC effects~\cite{Ethier:2020way}
are supposed to modify the gluon nPDF shape compared with the free proton case.
This might result in enhancements as large as 10\% in the \ttbar{} production cross-section compared with the same process
measured in \pp\ collisions.

The top quark, the heaviest elementary particle, is short-lived and decays through
$t\rightarrow W b$ with a branching ratio of almost 100\%.
The subsequent $W$ boson decay may proceed leptonically ($W\rightarrow \ell \nu_\ell,\;\ell=e,\mu$)
or hadronically ($W\rightarrow q\bar{q}^\prime$)~\cite{ParticleDataGroup:2022pth}.
At the LHC, top quarks are preferentially produced in
top quark--antiquark ($t\bar{t}$) pairs via gluon--gluon fusion and their production dominates
over single-top-quark production~\cite{Giammanco:2017xyn}.
With large integrated luminosities of \pPb\ and Pb+Pb datasets recorded between 2015 and 2018~(Run~2),
the observation of the $t\bar{t}$  process becomes accessible in HI collisions for the
first time at the LHC\@. In particular, $t\bar{t}$ events reconstructed in the
\ljets~($t\bar{t}\rightarrow WbW\bar{b}\rightarrow \bar{\ell}\nu_\ell bq\bar{q}^\prime\bar{b}$)
and dilepton~($t\bar{t}\rightarrow WbW\bar{b}\rightarrow \bar{\ell}\nu_\ell b\ell^\prime\bar{\nu_{\ell^\prime}}\bar{b}$) channels,
with relatively low expected background contributions, can be examined experimentally~\cite{dEnterria:2017jyt}.
The CMS experiment observed $t\bar{t}$ production using the \ljets{} decay channel in \pPb{} collisions at
$\sqrt{s_\mathrm{NN}}=8.16\;\mathrm{~\TeV}$~\cite{CMS:2017hnw}. Also, evidence of $t\bar{t}$ production
in Pb+Pb collisions at $\sqrt{s_\mathrm{NN}}=5.02\;\mathrm{~\TeV}$ was reported by the CMS Collaboration
with $4.0\sigma$ significance using the dilepton channel~\cite{CMS:2020aem}.

In this paper, a measurement of $t\bar{t}$ production in \pPb{} collision data collected at
$\sqrt{s_\mathrm{NN}}=8.16\;\mathrm{~\TeV}$ with the ATLAS experiment is presented.
Top-quark pairs are reconstructed in the \ljets{} and dilepton channels using final states with
electrons, muons and jets.
The dilepton mode is less abundant
than the \ljets\ channels but has a significantly higher purity.
The lower transverse-momentum requirements imposed on individual leptons and jets and
more precise detector calibration studies considered in the data analysis
lead to a significant improvement in the measurement precision compared with the result from Ref.~\cite{CMS:2017hnw}.
Also a nuclear modification factor for \ttbar\ production in \pPb\ is measured for the first time at the LHC.
The results are compared with calculations at next-to-next-to-leading-order~(NNLO)
in the strong coupling constant $\alpha_s$ involving the most up-to-date nPDF sets.

\section{ATLAS detector}
The ATLAS experiment~\cite{PERF-2007-01} at the LHC is a multipurpose particle detector
with a forward--backward symmetric cylindrical geometry and a near \(4\pi\) coverage in
solid angle.\footnote{
ATLAS uses a right-handed coordinate system with its origin at the nominal interaction point (IP)
in the centre of the detector and the \(z\)-axis along the beam pipe.
The \(x\)-axis points from the IP to the centre of the LHC ring,
and the \(y\)-axis points upwards.
Polar coordinates \((r,\phi)\) are used in the transverse plane,
\(\phi\) being the azimuthal angle around the \(z\)-axis.
The pseudorapidity is defined in terms of the polar angle \(\theta\) as \(\eta = -\ln \tan(\theta/2)\) and is equal to the rapidity
$ y = \frac{1}{2} \ln \left( \frac{E + p_z c}{E - p_z c} \right) $ in the relativistic limit.
Angular distance is measured in units of \(\Delta R \equiv \sqrt{(\Delta y)^{2} + (\Delta\phi)^{2}}\).
}
It consists of an inner tracking detector surrounded by a thin superconducting solenoid
providing a \qty{2}{\tesla} axial magnetic field, electromagnetic and hadronic calorimeters, and a muon spectrometer.
The inner tracking detector covers the pseudorapidity range \(|\eta| < 2.5\).
It consists of silicon pixel, silicon microstrip, and transition radiation tracking detectors.
Lead/liquid-argon (LAr) sampling calorimeters provide electromagnetic (EM) energy measurements
with high granularity within the region \(|\eta|< 3.2\).
A steel/scintillator-tile hadronic calorimeter covers the central pseudorapidity range (\(|\eta| < 1.7\)).
The endcap and forward regions are instrumented with LAr calorimeters
for EM and hadronic energy measurements up to \(|\eta| = 4.9\).
The muon spectrometer surrounds the calorimeters and is based on
three large superconducting air-core toroidal magnets with eight coils each.
The field integral of the toroids ranges between \num{2.0} and \qty{6.0}{\tesla\metre}
across most of the detector.
The muon spectrometer includes a system of precision tracking chambers up to \(|\eta| = 2.7\) and fast detectors for triggering up to \(|\eta| = 2.4\).
The luminosity is measured mainly by the LUCID--2~\cite{LUCID2} detector, which is located close to the beampipe.
A two-level trigger system is used to select events~\cite{TRIG-2016-01}.
The first-level trigger is implemented in hardware and uses a subset of the detector information
to accept events at a rate below \qty{100}{\kHz}.
This is followed by a software-based trigger that
reduces the accepted event rate to \qty{1}{\kHz} on average
depending on the data-taking conditions.
A software suite~\cite{ATL-SOFT-PUB-2021-001} is used in data simulation, in the reconstruction
and analysis of real and simulated data, in detector operations, and in the trigger and data acquisition
systems of the experiment.

\section{Data and simulated event samples}
The data used in this measurement were collected with the ATLAS detector during \pPb{} collisions
in 2016, and correspond to an integrated luminosity of 165~nb$^{-1}$.
The proton and Pb beams had an energy of 6.5~\TeV and 2.56~\TeV per nucleon, respectively,
resulting in a nucleon–nucleon centre-of-mass collision energy of 8.16~\TeV
and a rapidity boost of this frame of ±0.465 units relative to the ATLAS laboratory
frame, depending on the direction of the $p$ beam.
Two beam-direction configurations were provided with \pPb{} and \Pbp{} collisions with about a factor of two more
integrated luminosity in the latter configuration where the Pb beam goes in the $+z$ direction.
The average number of hadronic interactions per bunch crossing was 0.18.

Samples of Monte Carlo~(MC) simulated events are used to develop the analysis procedures, evaluate signal
and background contributions, estimate signal efficiencies, and provide predictions for comparison with data. All samples are
processed using the full ATLAS detector simulation~\cite{ATLAS:2010arf} based
on the \geant4 framework~\cite{GEANT4:2002zbu}.
All signal and simulated background samples are produced separately for two isospin
(proton--proton and proton-neutron) configurations and then embedded into real
\pPb{} or \Pbp{} data events for accurate UE modelling. These `data overlay
events' are then processed using the same reconstruction and analysis chain as the data.
Since the cross-section differences between the two isospin combinations are below 0.1\%,
rates of simulated events are scaled by the mass number $A_\mathrm{Pb}=208$ of the Pb nucleus
and a ratio of integrated luminosities in the data and MC simulation.
The top-quark mass is set to 172.5~\GeV in all top-quark samples.
The \evtgen{} program~\cite{Lange:2001uf} is used to treat the decays of $b$- and $c$-flavoured hadrons
in samples simulated using \powheg~\textsc{Box}~v2~\cite{Alioli:2010xd} and \madgraph{}~\cite{Alwall:2014hca} MC generators.

The nominal simulated $t\bar{t}$ sample is produced using the next-to-leading-order~(NLO) event generator
\powheg~\textsc{Box}~v2 with the NNPDF3.0NLO parton distribution function~(PDF) set~\cite{NNPDF:2014otw}. It is interfaced with
the \pythia~8.243~generator~\cite{Sjostrand:2014zea} using the NNPDF2.3LO PDF set~\cite{NNPDF:2012cx} and the A14 tune~\cite{ATL-PHYS-PUB-2014-021}
for the parton-shower and hadronisation modelling.
The \powheg{} resummation damping parameter $h_\mathrm{damp}$, which controls matrix element to parton-shower matching and effectively regulates the high-\pt{}  radiation, is set to 1.5 times the top-quark mass.
All signal samples are normalised using the NNLO+next-to-next-to-leading logarithmic~(NNLL) $t\bar{t}$ cross-section prediction from the \toppp~v2 program~\cite{Czakon:2011xx}.

Alternative $t\bar{t}$ simulation samples are generated to assess systematic uncertainties
related to the signal modelling.
One sample uses the \powheg~\textsc{Box}~v2 MC generator with the \herwig~v7.2 parton-shower
and hadronisation model~\cite{Bellm:2015jjp} employing the H7.2-Default tune~\cite{Bellm:2015jjp,Bellm:2019zci}.
Another sample is generated by employing the \madgraph{} generator in combination
with the NNPDF3.0NLO PDF set, while using the \pythia8 parton-shower and hadronisation model.
Uncertainties in the amount of parton-shower radiation are evaluated by generating \powheg~\textsc{Box}~v2+\pythia8 samples
with an increased cut-off scale for the first gluon emission, represented by the $h_\mathrm{damp}$ parameter,
which is set to three times the top-quark mass~\cite{ATL-PHYS-PUB-2016-020}.

The backgrounds that are evaluated with MC simulation arise from $W$ and $Z$ bosons produced
in association with jets, from single top-quark production and diboson production.
The $Z$+jets, and $W$+jets events are simulated with the \sherpa~v2.2.10 generator~\cite{Sherpa:2019gpd}
in combination with the NNPDF3.0NNLO PDF set~\cite{NNPDF:2014otw},
using NLO matrix elements for up to two partons, and LO matrix elements for up to four partons,
as discussed in Ref.~\cite{Hoche:2012tae}.
The $V$+jets~($V = W,Z$) samples are normalised to NNLO cross-sections~\cite{Anastasiou:2003ds}
and are further filtered for the content of light,
$c$- and $b$-quarks forming samples labelled as $W$+light/$Z$+light, $W$+$c$/$Z$+$c$ and $W$+$b$/$Z$+$b$ in the following.

The $t$-channel and $tW$ associated
production processes for single-top-quarks are simulated using the \powheg~\textsc{Box}~v2~\cite{Frederix:2012dh,Re:2010bp}
generator with the NNPDF3.04fNLO~\cite{NNPDF:2014otw} and NNPDF3.0NLO PDF sets, employing \pythia8 with the A14 tune
as the parton-shower and hadronisation model. The diagram removal scheme~\cite{Frixione:2008yi} is used
to treat the interference between the $t\bar{t}$ and $tW$ final states.
Smaller backgrounds from diboson production
($WW$, $WZ$, and $ZZ$) with additional jets are simulated using the \sherpa~v2.2.11  generator
with the NNPDF3.0NNLO PDF set.

\section{Event selection and background estimation}

The events are selected using single-lepton electron or muon triggers with a minimum
transverse momentum ($\pT$) threshold of 15~\GeV~\cite{ATLAS:2019dpa,ATLAS:2020gty}.
They are required to have at least one reconstructed
vertex built from at least two good-quality charged-particle tracks with $\pT > 0.1\;\mathrm{~\GeV}$.

Electron candidates are reconstructed from a localised cluster of energy deposits in
the EM calorimeter matched to a track in the ID\@. They are further required to satisfy the `Medium' likelihood-based
requirements~\cite{ATLAS:2019qmc} and to have
$\pt > 18\;\mathrm{~\GeV}$ and $|\eta_\mathrm{cluster}| < 2.47$.
The electron candidates are required to be isolated using varying requirements on track- and calorimeter-based isolation~\cite{ATLAS:2019qmc}.

Muon candidates are reconstructed by combining matching tracks reconstructed in the inner detector
and the muon spectrometer, and are required to satisfy the `Medium' quality requirements~\cite{ATLAS:2020auj}.
The muon candidates are also required to have $\pt > 18\;\mathrm{~\GeV}$ and $|\eta| < 2.5$,
and to be isolated using fixed criteria for calorimeter-based isolation and varying requirements on track-based isolation~\cite{ATLAS:2020auj}.
Lepton tracks have to fulfil further requirements described in Ref.~\cite{ATLAS:2022jbj}.

Jets are reconstructed from calorimeter energy deposits~\cite{ATLAS:2019jgo},
using the anti-$k_t$ algorithm~\cite{Cacciari:2008gp,Cacciari:2011ma} with a radius parameter $R = 0.4$.
The jet kinematics are corrected event-by-event for the contribution from underlying event~(UE),
and are calibrated using simulations of the
calorimeter response and in situ measurements of the absolute energy scale~\cite{ATLAS:2020cli}.
In-situ measurements are carried out in $pp$ collisions and cross-calibrated to the \pPb\ system.
Jets reconstructed this way are referred to as HI jets. The kinematic variables are calculated using these jets.
However, the $b$-tagging information is not available for HI jets.
For this reason, a second type of jets is also used in the analysis.

The second type of jets is reconstructed from particle-flow~(PF) objects that combine information from
topological clusters of calorimeter energy deposits and charged-particle tracks~\cite{ATLAS:2017ghe}. The PF jets are
built with the anti-$k_t$ algorithm with a radius parameter $R = 0.4$ and calibrated in the same way as in high pile-up
$pp$ collisions at $\sqrt{s}= 13\;\mathrm{TeV}$~\cite{FTAG-2019-07}.
PF jets containing $b$-hadrons are tagged using the DL1r algorithm~\cite{ATLAS:2019bwq},
a multivariate discriminant based on deep-learning techniques making use of track impact parameters
and reconstructed secondary vertices. A tagger working point with 85\% efficiency
(evaluated in simulated $t\bar{t}$ events) in $pp$ collisions for tagging $b$-quark
jets from top-quark decays is used, corresponding to rejection factors
of about three against $c$-quark jets and 40 against light-quark and gluon jets.

HI jets with $\pt>20\;\mathrm{GeV}$ and $|\eta| < 2.5$ are matched to PF jets in the same event using
a geometric criterion of minimal $\Delta R$. The $b$-tagging information
is inherited from matched PF jets if $\Delta R<0.3$
between a HI and a PF jet. HI jets lacking a PF counterpart are considered as non-$b$-tagged.
This mismatch arises predominantly from differences in energy
calibration as HI jets have a dedicated calibration optimised for the low pile-up environment.

To prevent the double-counting of electron energy deposits as jets, the closest jet to an electron
candidate is removed if it is within $\Delta R \leq 0.2$ of the electron.
Furthermore, to reduce the contribution of leptons from heavy-flavour hadron decays inside jets,
leptons within $\Delta R \leq 0.4$ of selected jets are discarded, unless the lepton is a muon and
the jet has fewer than three associated tracks, in which case the jet is discarded. This approach is
applied separately to both jet types prior to the HI-PF jet matching.

The missing transverse momentum, with magnitude \Etmiss, is computed using momenta of fully calibrated leptons, photons, and PF jets,
combined with the soft hadronic activity measured by reconstructed charged-particle
tracks not associated with the hard objects~\cite{PERF-2016-07}.
No requirement on \Etmiss\ is imposed in the final event selections used for this measurement.

Events with exactly two opposite-sign leptons form the dilepton channel.
Events with same-flavour lepton pairs ($e^+e^-$ or $\mu^+\mu^-$) with the invariant mass ($m_{\ell\ell}$) within a $Z$ boson mass
window~($80<m_{\ell\ell}<100\;\mathrm{GeV}$) are discarded, and $m_{\ell\ell}$ must be greater than
15 (45)~\GeV in the $e\mu$ ($ee$ and $\mu\mu$) channel. The latter condition is imposed to match the phase
space of the $Z$+jets simulation sample and does not have a significant impact on the results.
Events are further required to have at least two HI jets including at least one $b$-tagged jet.
Such events form the signal region~(SR) of the dilepton channel.
The regions with two leptons and exactly one or at least two $b$-tagged jets
are labelled as $2\ell1b$ and $2\ell2b\mathrm{incl}$, respectively.

Events with exactly one lepton and at least four HI jets including at least one $b$-tagged jet form the SR of the \ljets\ channel.
Based on the lepton flavour, the \ljets{} SR is further split into four regions with
one electron or muon and exactly one or at least two $b$-tagged jets,
labelled as $1\ell1b$~$e$+jets, $1\ell2b\mathrm{incl}$~$e$+jets, $1\ell1b$~$\mu$+jets and $1\ell2b\mathrm{incl}$~$\mu$+jets.

Non-prompt leptons, hadrons and photons that meet the lepton selection criteria are sources of the non-prompt and misidentified lepton background,
commonly referred to as fake-lepton background. In this measurement, the normalisation and shape of the fake-lepton background are
estimated from data using a technique called the Matrix Method~(MM)~\cite{ATLAS:2022swp}.
The MM technique exploits differences for lepton-identification-related characteristics between prompt,
isolated leptons originating from $W$ and $Z$ boson decays (referred to as real leptons), and leptons that are either non-isolated or result from the misidentification of photons or jets (referred to as fake leptons).
The method expresses the number of selected events in each sample as a linear combination of the numbers of events with real and fake leptons, where the coefficients are determined using ‘Tight’ and ‘Loose’ lepton selections and are related to the probabilities (efficiencies) for ‘Loose’ leptons to also satisfy the ‘Tight’ selection criteria.
`Tight' and `Loose' selections are used to measure both the real- and fake-lepton efficiencies. The `Tight' selection employs the complete set of high-quality lepton identification criteria. The `Loose' lepton selection is obtained from
the `Tight' selection by relaxing the identification requirements and dropping the lepton isolation requirements~\cite{ATLAS:2019qmc,ATLAS:2020auj}.
The real-lepton efficiencies are evaluated in MC samples ($Z$+jets, $\ttbar{}$) as the ratio of the numbers of
simulated prompt leptons passing the `Tight' requirements and simulated prompt leptons passing the `Loose' requirements. The fake-lepton efficiencies
are estimated using dedicated control regions~(CR) with one lepton passing `Loose' identification and isolation requirements,
and missing transverse energy $\met<20\;\mathrm{GeV}$. Due to limited statistical precision in the CR with two leptons, the one-lepton CR is
also used to evaluate the fake-lepton contribution for the SR in the dilepton channel.
Both the real- and fake-lepton efficiencies are evaluated for electrons and muons in $\pt$ and $|\eta|$ bins in event categories
with zero, one and at least two $b$-tagged jets.

Fake-lepton efficiencies vary between 15\%--22\% for electrons and 0.5\%--10\% for muons, and are larger at lower lepton \pt.
Data events which satisfy the baseline analysis selection imposed on `Loose' leptons
are weighted according to the efficiencies for both the prompt and fake leptons.
To validate the method, the predictions are compared with data in a CR with a~larger
fraction of fake-lepton candidates than expected in the analysis SR
and a satisfactory agreement is found.
This CR, which is defined by imposing the dilepton and \ljets{} selection and requiring no $b$-tagged jets,
is dominated by $W$/$Z$+jets processes.

The main background contribution in the dilepton channel includes the $Z$+jets and single-top $tW$ processes.
Up to 5\% of $Z$+jets events originate from the $Z\rightarrow \tau^+\tau^-$ process.
This fraction rises to about 99\% in selected events in the $e\mu$ channel.
The background in \ljets{} channel is formed mainly by $W$+jets events and the fake-lepton contribution.
The expected signal fractions in the SRs are 21\% and 73\% in the \ljets{} $1b$ and $\geq 2b$ regions, respectively;
and 53\%, and 91\% in the dilepton 1b and $\geq 2b$ regions, respectively.

\section{Systematic uncertainties}
\label{sec:syst}
Systematic uncertainties affecting the measurement arise from the reconstruction of leptons and jets,
$b$-tagging, fake-lepton background, the signal and background modelling, and integrated luminosity.

Uncertainties in the muon momentum scale and resolution follow those in Ref.~\cite{ATLAS:2020auj}.
The analysis includes uncertainties in the data-to-MC correction factors applied to simulated samples
for the muon reconstruction, isolation, track-to-vertex-association, and the trigger efficiencies,
evaluated using $Z\rightarrow \mu^+\mu^-$ events in \pPb\ collisions.
Uncertainties in the electron reconstruction, identification, isolation and trigger
are derived using $Z\rightarrow e^+e^-$ events
in \pPb\ collisions and the uncertainty in the low pile-up
energy calibration is evaluated in accordance with Ref.~\cite{ATLAS:2019qmc}.
The jet-related uncertainties are derived from in situ studies of the calorimeter
response~\cite{ATLAS:2020cli} and their application to the jets used in
HI data~\cite{ATLAS-CONF-2015-016},
and from comparisons of the simulated response in samples from different generators.
The $b$-tagging systematic uncertainties are
computed by varying the data-to-MC correction factors within their
uncertainties~\cite{FTAG-2018-01,FTAG-2020-08,FTAG-2019-02}. To assess uncertainties in the HI-PF jet matching,
two systematic variations are introduced. The first uncertainty adjusts the matching distance  $\Delta R$ to $\pm 0.1$ from the default
$\Delta R = 0.3$ for matching $b$-tagged jets. The second accounts for events where HI jets lack a
PF counterpart (18\% of jets in data and 15\% in the signal and background
MC samples), in which case HI jets are randomly considered as $b$-tagged based on the light-flavour
jet mistag rate~\cite{FTAG-2019-02}.
The resulting systematic variation has a negligible effect on the final result.

Additional systematic uncertainties related to the normalisation of the $V$+jets samples
are determined using the Berends scaling technique~\cite{Berger:2010zx,Berends:1990ax,Ellis:1985vn}. Single-top-quark
diagram removal and diagram subtraction variation samples are used to assess the uncertainties from the
interference between the \ttbar{} and $tW$ processes~\cite{Re:2010bp}. A conservative uncertainty of $9.5$\% is
considered for the normalisation of both the $tW$ and $t$-channel single-top-quark processes~\cite{ATLAS:2022jbj}.
The diboson background normalisation is allowed to vary by 50\%~\cite{ATLAS:2023slx}.

Systematic uncertainties of the fake-lepton background estimate in both \ljets{} and dilepton channels arise from statistical and systematic variations of the real- and fake-lepton efficiencies, and are evaluated using the MM technique.
Additional conservative uncertainties of
100\% of the normalisation in the \mujets\ and 50\% of the normalisation in the \ejets\ and the dilepton SRs are imposed as uncorrelated uncertainties.
The magnitudes of the systematic variations in the normalisation of the fake-lepton background
are inferred from the agreement of the data to prediction in a zero $b$-tagged jets~($0b$) CR. Additional shape variations of this background in the \ljets{} channel are evaluated also in the $0b$ CR\@.
To derive the fake-lepton background shape variations, all background contributions except the fake-lepton events are subtracted from the data and the difference is normalised to the number of fake-lepton events.
A ratio is constructed of such subtracted and scaled data to the fake-lepton contribution
as a function of the azimuthal angle $\Delta\phi(\Etmiss,\ell)$ between the lepton and \met.
Values of this ratio vary from 0.5 to 3.5 in bins of $\Delta\phi(\Etmiss,\ell)$. The ratio is fit by a second-order polynomial.
Shape variations of the fake-lepton background in $1b$ and $\geq 2b$ \ljets{} SRs
are defined as up and down fit shape variations using the fit parameter uncertainties.

The shape of the $\Delta\phi(\Etmiss,\ell)$ variable is not correlated to the shape of the fit variable described in Section~\ref{sec:results}.

Uncertainties due to the choice of the parton-shower and hadronisation models in addition to  the matrix-element matching to the parton shower
are estimated by using the alternative \ttbar{} MC samples.
The uncertainty due to initial-state radiation~(ISR) is estimated by variations of $\alpha_s$ for ISR in the A14 tune~\cite{ATL-PHYS-PUB-2014-021}.
Further effects on the ISR are evaluated by varying the renormalisation~($\mu_\mathrm{r}$) and factorisation scales~($\mu_\mathrm{f}$)
in the matrix-element calculation as well as the $h_\mathrm{damp}$ parameter.
The $\mu_\mathrm{r}$ and $\mu_\mathrm{f}$ are varied independently by factors of 0.5 and 2.0 avoiding same side variations of the scales.
The effect of final-state radiation~(FSR) uncertainties is evaluated by modifying the
$\mu_\mathrm{r}$ for emissions from the parton shower by factors of 0.5 and 2.0.
The PDF uncertainties affecting the \ttbar\ signal are evaluated using the PDF4LHC15
Hessian uncertainties~\cite{Butterworth:2015oua}.

The uncertainty in the integrated luminosity of the combined data sample is 2.4\%.
It is derived from the calibration of the luminosity scale using $x$-$y$ beam-separation scans,
following a methodology similar to that detailed in Ref.~\cite{ATLAS:2022hro}, and using the LUCID-2
detector for the baseline luminosity measurements~\cite{LUCID2}.

\section{Results}
\label{sec:results}
The signal strength \mutt, defined as the ratio of the observed signal for the combined \ljets{} and dilepton final states
to the SM expectation with no nPDF effects included, is measured using a binned profile-likelihood method~\cite{Cranmer:1456844}.
The parameter \mutt{} is determined by the fit to the \Ht{} data distributions in the six SRs,
where the \Ht{} variable is defined as the scalar sum of the transverse momenta of the leptons and HI jets.
In the fit, systematic uncertainties are represented by nuisance parameters, which are additional fit parameters constrained by a Gaussian-distributed probability density.
By allowing the nuisance parameters to shift from their expected values of zero, the best global fit to the data is achieved.
This procedure permits an improved description of the data by combined signal and background contributions, considering their modelling in terms of shapes
and normalisation, and the effects of experimental uncertainties, which leads to a reduction of the total systematic uncertainty in the parameter of interest.
The \Ht{} distributions predicted by the fit are shown in Figure~\ref{fig:data_postfit_full} for the six SRs.
Distributions predicted by the fit and the observed distributions are in reasonable agreement.

\begin{DIFnomarkup}
\begin{figure}[!htb]
\centering
\subfigure[]{\includegraphics[width=0.32\textwidth]{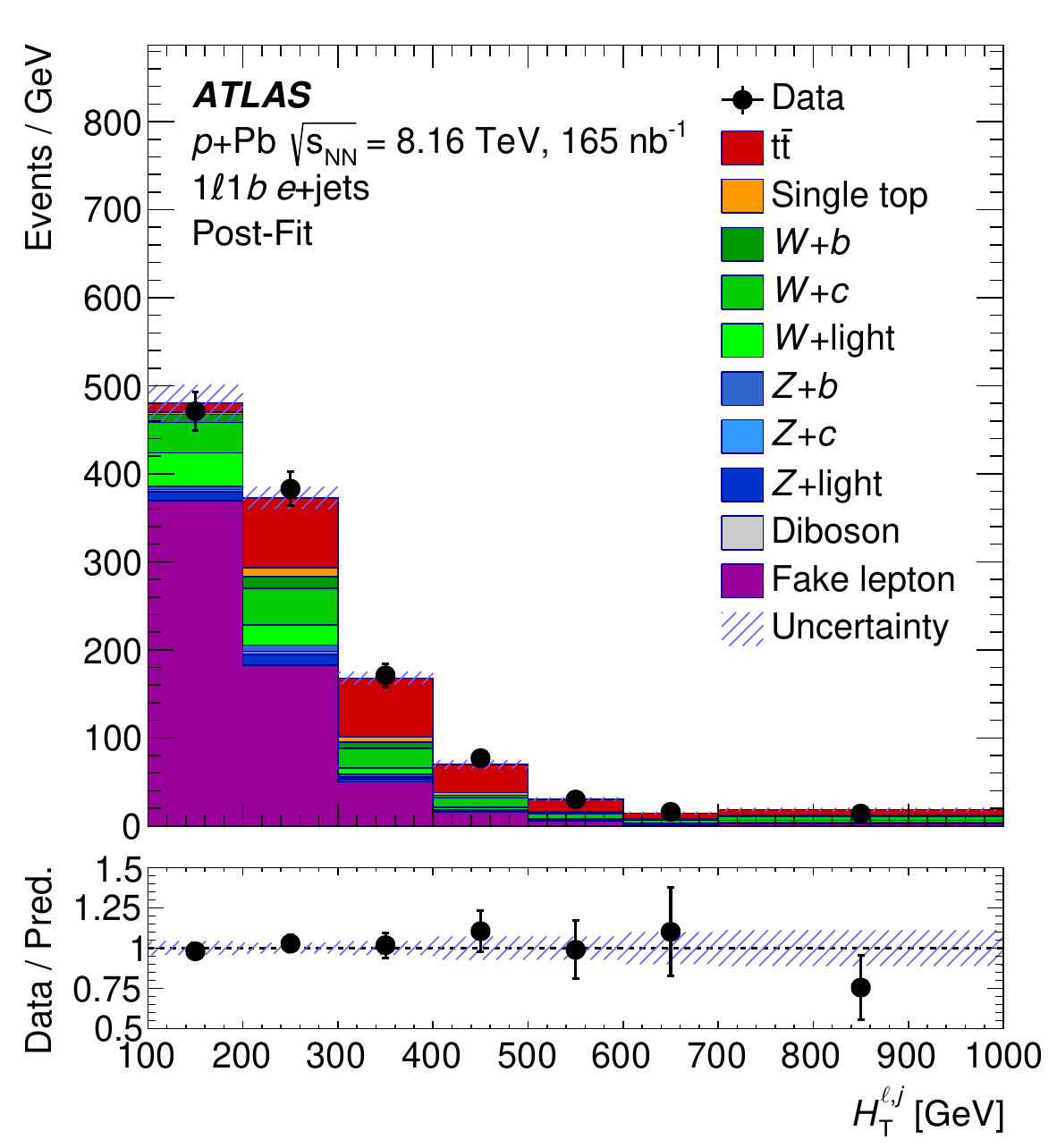}}
\subfigure[]{\includegraphics[width=0.32\textwidth]{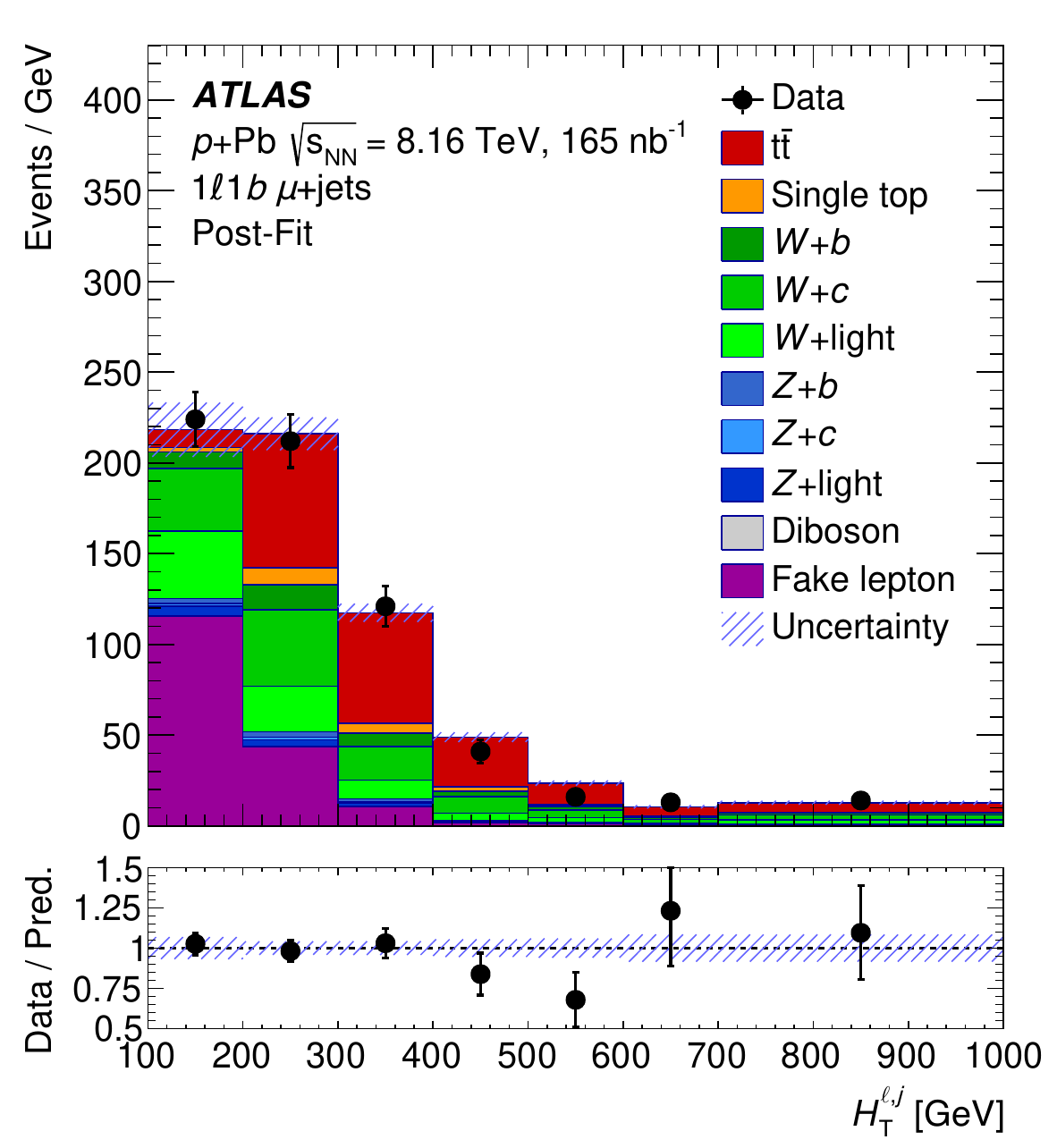}}
\subfigure[]{\includegraphics[width=0.32\textwidth]{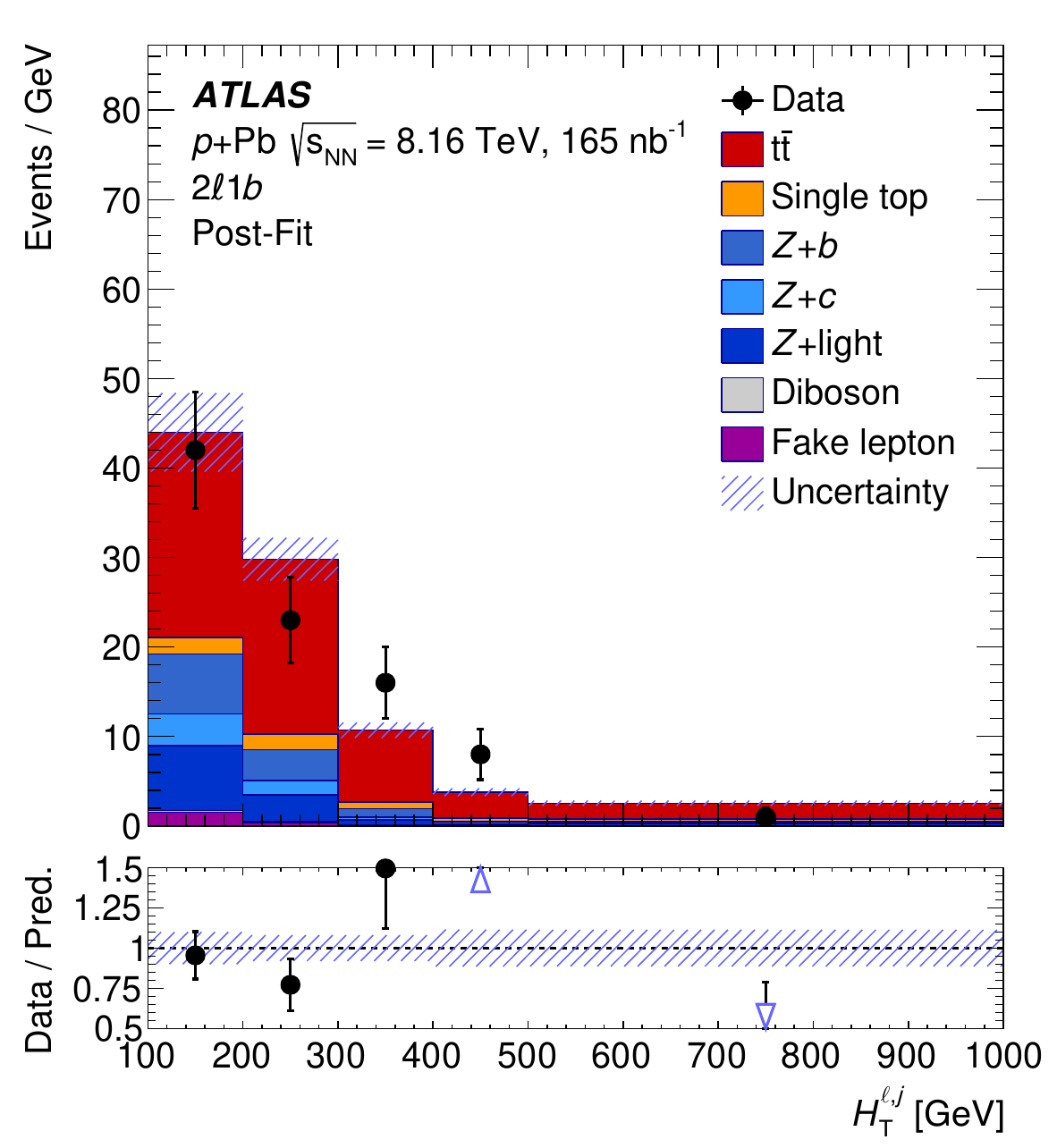}} \\
\subfigure[]{\includegraphics[width=0.32\textwidth]{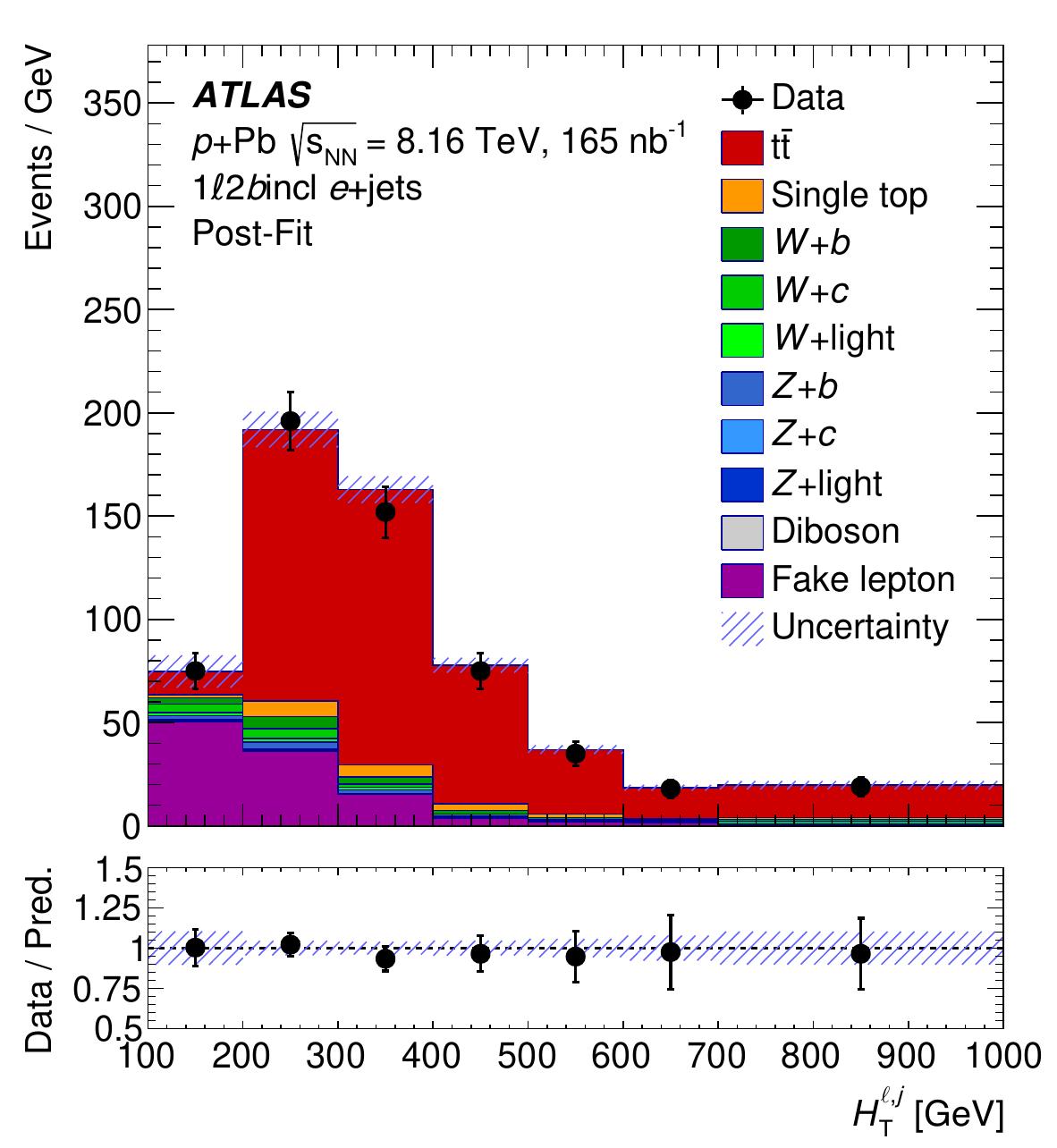}}
\subfigure[]{\includegraphics[width=0.32\textwidth]{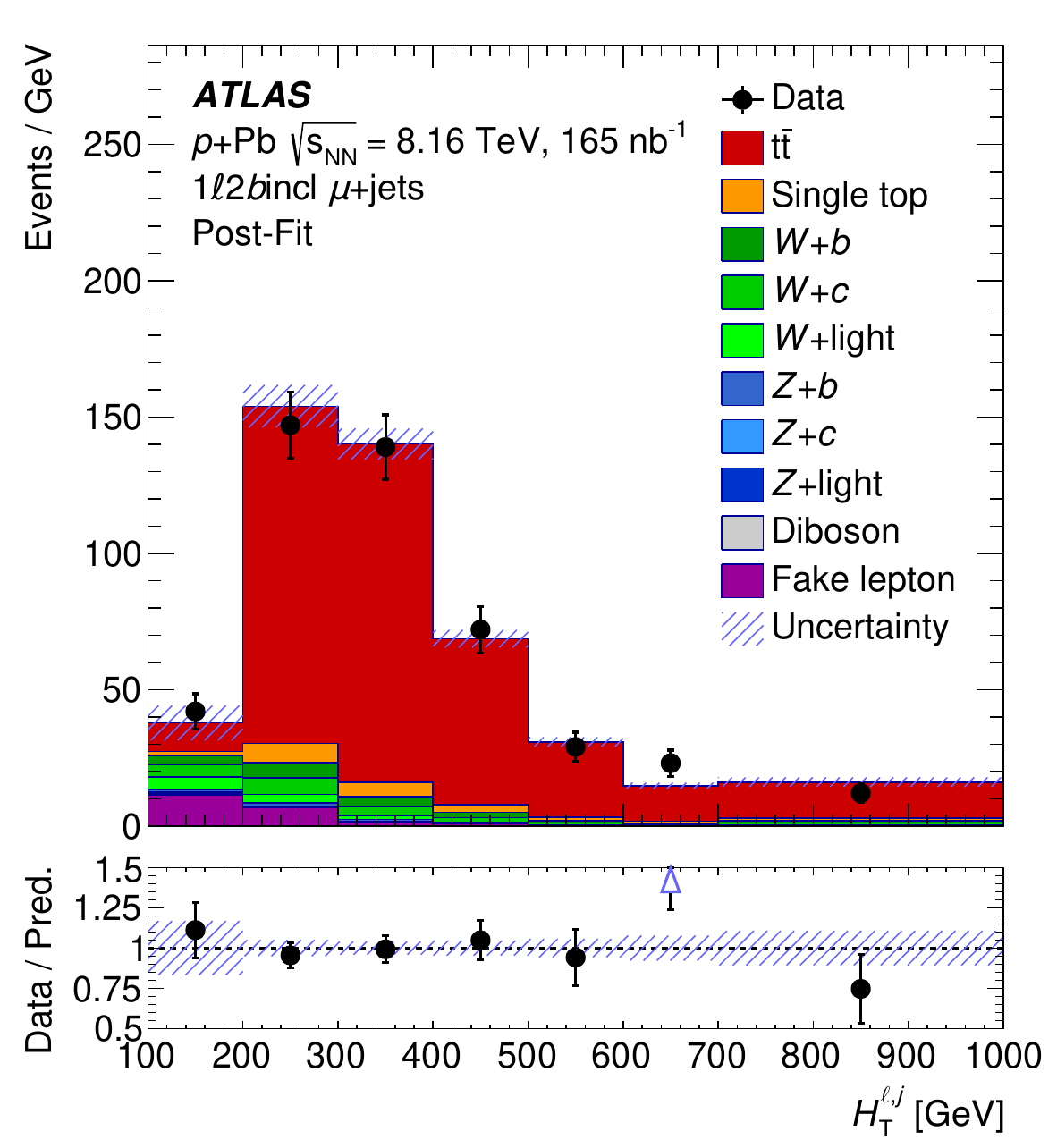}}
\subfigure[]{\includegraphics[width=0.32\textwidth]{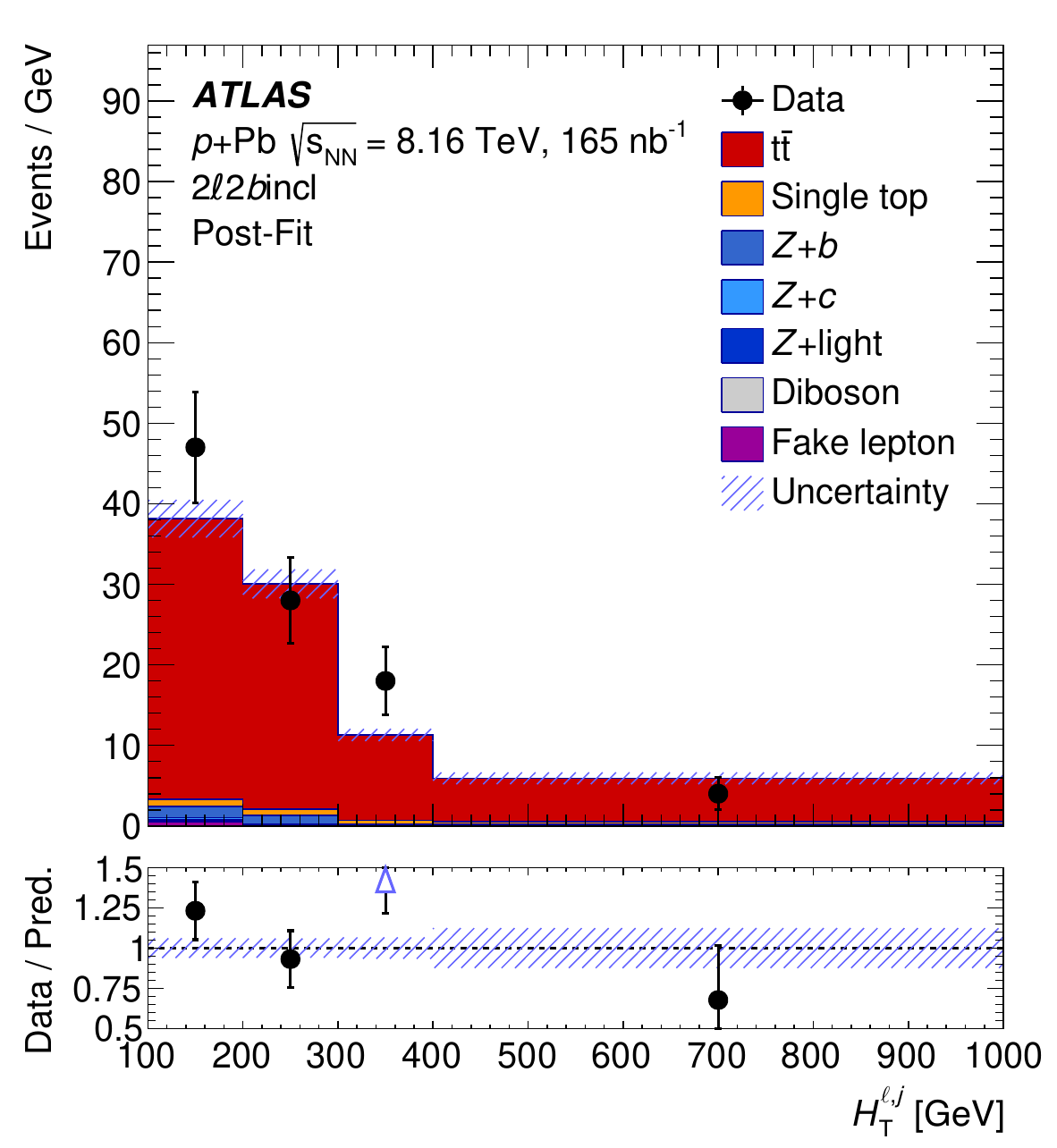}}
\caption{Comparison of data and total post-fit prediction for the \Ht\ distribution in each of the six SRs ($e$+jets: (a)~$1\ell1b$ and
(d)~$1\ell2b\mathrm{incl}$, $\mu$+jets: (b)~$1\ell1b$ and (e)~$1\ell2b\mathrm{incl}$, dilepton: (c)~$2\ell1b$ and
(f)~$2\ell2b\mathrm{incl}$), with uncertainties represented by the hatched area. The full markers in the
bottom panels show a ratio of data and a sum of predictions. Open triangles indicate bins with
entries which are outside the ratio range. The first and last bins include underflow and overflow events, respectively.
The vertical order of the individual contributions forming the total prediction is the same as in the legend.
The $Z$+jets contribution is negligible in the \ljets{} $\geq 2b$ regions.}
\label{fig:data_postfit_full}
\end{figure}
\end{DIFnomarkup}

All systematic uncertainties in the fit are treated as correlated over the SRs
unless stated otherwise in Section~\ref{sec:syst}.
The uncertainties associated with the fake-lepton background are notably constrained during the fitting process.
The leading contributions to the total systematic uncertainty are the
jet energy scale and signal modelling.
Table~\ref{tab:syst_group_data} shows a breakdown of relative systematic uncertainties on the cross-section in data.
The total relative systematic uncertainty amounts to 8\%.

\begin{DIFnomarkup}
\begin{table}[!htb]
\centering
\caption{Summary of the impact of the systematic uncertainties on \ttbar\ cross-section grouped into different categories. The quoted uncertainties are obtained by repeating the fit with a group of nuisance parameters fixed to their fitted values and subtracting in quadrature the resulting total uncertainty from the uncertainty of the complete fit. However, the total uncertainty is not the quadratic sum of the grouped impacts, as this approach neglects the correlation among the different groups.}

\begin{tabular}{l|cc}
\hline
\multirow{2}{*}{Source} 		& \multicolumn{2}{c}{$\Delta\sigma_{\ttbar} / \sigma_{\ttbar}$} \\
&unc. up [\%]	&unc. down [\%]	\\
\hline
Jet energy scale					& +4.6				&-4.1					\\
$t\bar{t}$ generator				& +4.5				&-4.0					\\
Fake-lepton background		& +3.1				&-2.8					\\
Background							& +3.1				&-2.6					\\
Luminosity							& +2.8				&-2.5					\\
Muon uncertainties				& +2.3				&-2.0					\\
$W+$jets								& +2.2				&-2.0					\\
$b$-tagging						& +2.1				&-1.9					\\
Electron uncertainties			& +1.8				&-1.5					\\
MC statistical uncertainties	& +1.1				&-1.0					\\
Jet energy resolution			& +0.4				&-0.4					\\
$t\bar{t}$ PDF						& +0.1				&-0.1					\\
\hline
\hline
Systematic uncertainty			& +8.3				&-7.6					\\
\hline
\end{tabular}


\label{tab:syst_group_data}
\end{table}
\end{DIFnomarkup}

The background-only hypothesis is rejected with a significance of more than five standard deviations,
establishing the observation of the $t\bar{t}$ process in \pPb{} collisions by ATLAS.
Figure~\ref{fig:mu} presents the signal strength \mutt\ obtained in each region separately
and in the combined fit.
The fitted \mutt{} values in individual channels are consistent within uncertainties and within the SM prediction.
The precision of the \mutt{} value is limited by systematic uncertainties in the \ljets\ SRs while
the statistical uncertainties dominate in the dilepton SRs.
The significance is extracted using separate fits of \mutt{} to the combined four \ljets{}
and combined two dilepton SRs, and exceeds in both cases by five standard deviations.
This establishes the observation of \ttbar\ production in the individual \ljets\ and dilepton channels.
The latter is reported for the first time in \pPb\ collisions at the LHC.

The measured \mutt{} value is translated to the inclusive $t\bar{t}$ production cross-section~(\xsectt)
using the formula:\\
\begin{equation}
\sigma_{\ttbar}=  \mu_{\ttbar}\cdot A_\mathrm{Pb}\cdot \sigma^\mathrm{th}_{\ttbar},
\end{equation}
where $A_\mathrm{Pb}=208$ is the lead mass number and $\sigma^\mathrm{th}_{\ttbar}$ %
is the predicted $t\bar{t}$ production cross-section in nucleon-nucleon collisions derived
at the NNLO precision used to normalise the signal $\ttbar$ samples in \ljets{} and dilepton decay
modes~\cite{Alioli:2010xd,Czakon:2011xx}.
The measured inclusive $\ttbar{}$ cross-section for \pPb\ collisions is
$\sigma_{t\bar{t}}= 58.1\pm 2.0\;\mathrm{(stat.)}\;^{+4.8}_{-4.4} \;\mathrm{(syst.)}\;\mathrm{nb}
= 58.1\;^{+5.2}_{-4.9} \;\mathrm{(tot.)}\;\mathrm{nb}$.
The combined relative uncertainty amounts to 9\% and is dominated by the systematic contribution.

\begin{DIFnomarkup}
\begin{figure}[!htb]
\centering
\includegraphics[width=0.75\textwidth]{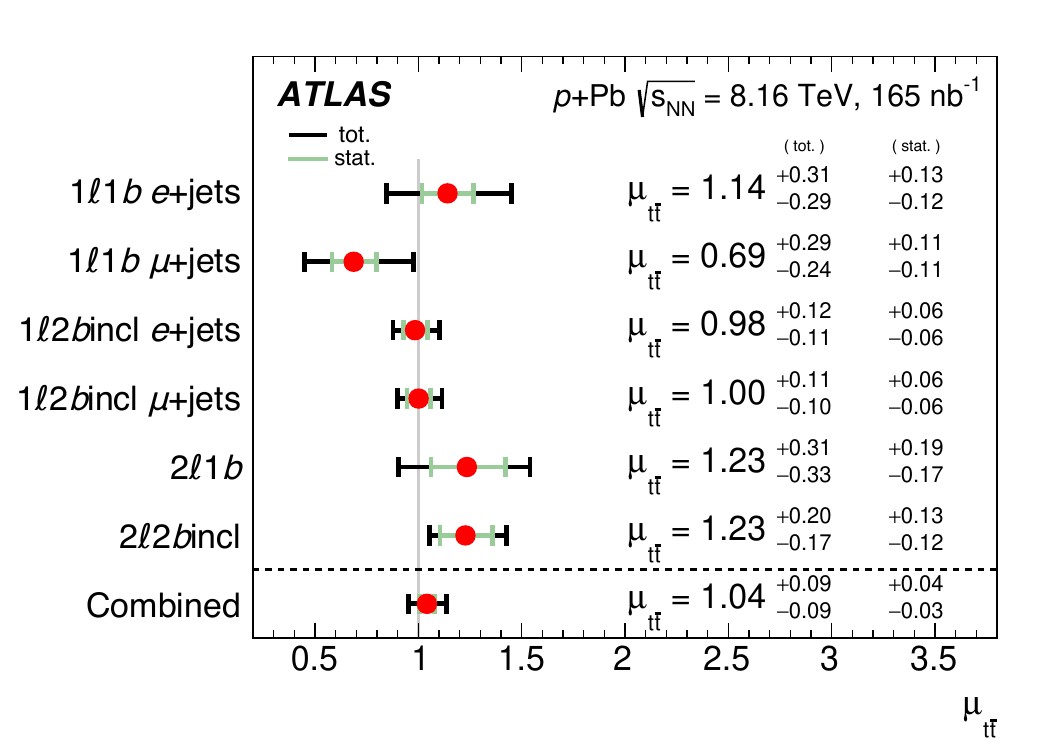}
\caption{The observed best-fit values of the signal strength $\mu_{t\bar{t}}$ and their uncertainties by final-state category and combined. The individual $\mu_{t\bar{t}}$ values for the channels are obtained from a simultaneous fit with the signal-strength parameter for each channel floating independently. The SM prediction is $\mu_{t\bar{t}} = 1$.}
\label{fig:mu}
\end{figure}
\end{DIFnomarkup}

Figure~\ref{fig:xsec} shows a comparison of the observed \xsectt{} with the measurement by CMS in \pPb\ collisions at
$\sqrt{s_\mathrm{NN}}=8.16\;\mathrm{~\TeV}$~\cite{CMS:2017hnw}.
The two results are in agreement within 1.4 standard deviations.
Also the most precise measurement of the $t\bar{t}$ production cross-section in \pp\ collisions at
$\sqrt{s}=8\;\mathrm{~\TeV}$ from the ATLAS and CMS combination~\cite{ATLAS:2022aof} is shown.
The cross-section value is extrapolated to the centre-of-mass energy of this measurement using the \toppp~v2 prediction
and scaled by $A_\mathrm{Pb}$ to the \pPb\ system. The extrapolated cross-section has
a 2.5\% relative uncertainty and does not involve any dependence on nPDF\@.
The extrapolation factor amounts to $1.0528 \pm 0.0005\;\mathrm{(PDF)}\;^{+0.0001}_{-0.0013}\;\mathrm{(scale)}$.
The measured cross-section is also compared with NLO calculations obtained with the \mcfm\
generator~\cite{Campbell:2016dks} scaled to the NNLO precision in QCD using the $K$-factor ($K=1.139$)
derived using the \toppp~v2 generator.
Four nPDF sets are used as input to the \mcfm\ calculations: EPPS21~\cite{Eskola:2021mjl},
nCTEQ15HQ~\cite{Kovarik:2015cma,Duwentaster:2022kpv}, nNNPDF30~\cite{Ubiali:2014bva,AbdulKhalek:2022fyi}
and TUJU21~\cite{Helenius:2022rmf}.
The largest discrepancy is found for the nNNPDF30 nPDF set which does not include the recent Run~2 LHC data for heavy-flavour
production from \pPb\ collisions~\cite{Klasen:2023uqj}.
The remaining nPDF sets are in good agreement with the measured cross-section value.

A nuclear modification factor defined as
\begin{equation}
R_{p\mathrm{A}} = \frac{ \sigma_{t\bar{t}}^{p+\mathrm{Pb}} }{ A_\mathrm{Pb} \cdot \sigma_{t\bar{t}}^{pp} }
\end{equation}
is extracted using the measured \ttbar\ cross-sections
in \pPb\ collisions at $\sqrt{s_\mathrm{NN}}=8.16$~\TeV and \pp\ collisions at
$\sqrt{s}=8$~\TeV~\cite{ATLAS:2022aof}.
The latter is extrapolated to the centre-of-mass energy of the \pPb\ system.
All uncertainties are assumed to be uncorrelated in the cross-section measurements in \pPb\
and \pp.

The nuclear modification factor is measured to be $R_{p\mathrm{A}} = 1.090\pm0.039\;\mathrm{(stat.)}\;^{+0.094}_{-0.087}\;\mathrm{(syst.)} = 1.090\pm0.100\;\mathrm{(tot.)}$.
A comparison between the measured $R_{p\mathrm{A}}$ in data and theory is shown in Figure~\ref{fig:RpA}.
The measured value is found to be consistent with unity within the uncertainty.

\RpA\ is also calculated at NNLO precision using the \mcfm\ code~\cite{Campbell:2016dks}
scaled to the \pPb\ system for four different nPDF sets.
The uncertainty associated with the baseline PDF for \pp\ interactions is considered fully correlated in the predictions
and cancels out in the ratio. The resulting uncertainty represents the
uncertainty on nPDF\@. All nPDF calculations result in \RpA\ values above unity.
A good agreement is found between the measured and predicted \RpA.
The largest difference of more than one standard deviation above the measured \RpA\ value
is observed for the nNNPDF30 prediction.

\begin{DIFnomarkup}
\begin{figure}[!htb]
\centering
\subfigure[]{
\includegraphics[width=0.48\textwidth]{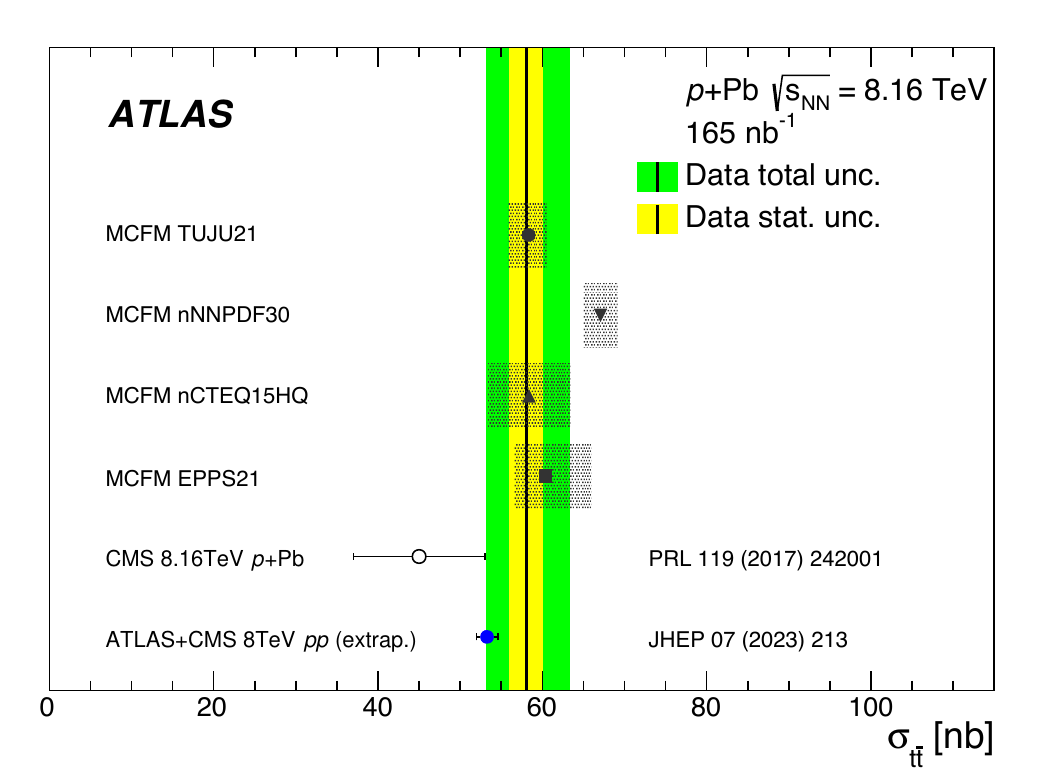}
\label{fig:xsec}}
\subfigure[]{
\includegraphics[width=0.48\textwidth]{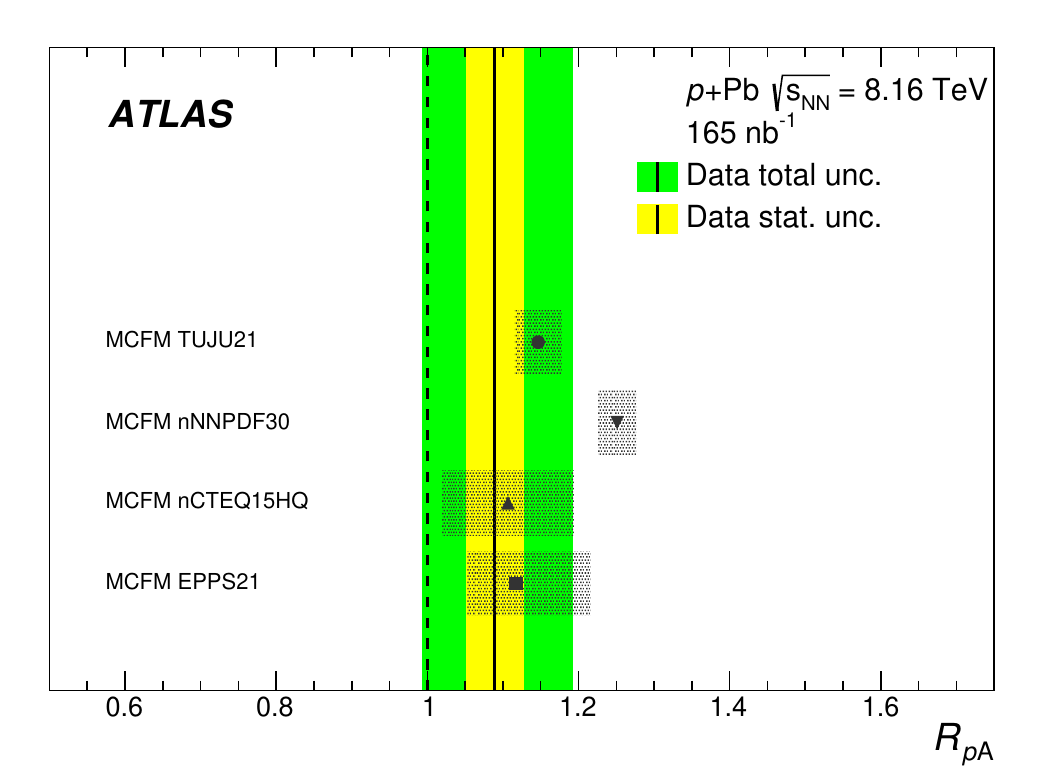}
\label{fig:RpA}}
\caption{Comparison between measured and predicted values of (a)~\xsectt{} and (b)~\RpA.
\xsectt{} is also compared with the existing measurement in \pPb\ collisions at $\sqrt{s_\mathrm{NN}}=8.16\;\mathrm{~\TeV}$~\cite{CMS:2017hnw},
and the combined measurement of $t\bar{t}$ production cross-section in \pp\ collisions at
$\sqrt{s}=8\;\mathrm{~\TeV}$ from ATLAS and CMS collaborations~\cite{ATLAS:2022aof}.
The latter is extrapolated to the centre-of-mass energy of this measurement and
is using the $A_\mathrm{Pb}$ factor.
Predictions are calculated at NNLO precision using the \mcfm\ code~\cite{Campbell:2016dks} scaled
to the \pPb\ system and given for different nPDF sets.
The uncertainty in the predictions represents the internal PDF uncertainty.
The solid black line indicates the measured value.
The combined statistical and systematic uncertainty of the measurement is represented by the outer
band around the central value, while the statistical component is depicted as the inner band.
}
\end{figure}
\end{DIFnomarkup}

\FloatBarrier
\section{Conclusion}
This paper reports a measurement of top-quark pair production in \pPb\ collisions
at the centre-of-mass energy $\sqrt{s_\mathrm{NN}}=8.16\;\mathrm{~\TeV}$ per nucleon pair with the ATLAS
experiment. Top-quark pairs are observed in the individual \ljets{} and dilepton channels
with electrons and muons in the final state.
The top-quark pair production in the dilepton channel is observed with significance exceeding
five standard deviations for the first time in the \pPb\ system at the LHC\@.
From the combination of both channels, the cross-section is measured with a relative uncertainty of 9\%,
which makes this measurement the most precise \ttbar\ cross-section determination in nuclear
collisions to date.
The measured cross-section is found to be in good agreement with a previous
measurement by the CMS Collaboration and with SM predictions.
A measurement of the nuclear modification factor is
reported using an extrapolation of the previously measured cross-section in \pp\
collisions at $\sqrt{s}=8\;\mathrm{~\TeV}$, based on a perturbative QCD calculation at NNLO.
Good agreement is found between the measured and
predicted \RpA\ values involving most of the state-of-the-art nPDF sets.
The largest deviation, of more than one standard deviation, is found for the nNNPDF30 set.
This measurement paves a new way to constrain nPDFs in the high Bjorken-$x$ region.
As such it is also an important input for upcoming measurements involving the extraction of QGP properties in
Pb+Pb collisions at the LHC.

\section*{Acknowledgements}
%

%
%

%
%

We thank CERN for the very successful operation of the LHC and its injectors, as well as the support staff at
CERN and at our institutions worldwide without whom ATLAS could not be operated efficiently.

The crucial computing support from all WLCG partners is acknowledged gratefully, in particular from CERN, the ATLAS Tier-1 facilities at TRIUMF/SFU (Canada), NDGF (Denmark, Norway, Sweden), CC-IN2P3 (France), KIT/GridKA (Germany), INFN-CNAF (Italy), NL-T1 (Netherlands), PIC (Spain), RAL (UK) and BNL (USA), the Tier-2 facilities worldwide and large non-WLCG resource providers. Major contributors of computing resources are listed in Ref.~\cite{ATL-SOFT-PUB-2023-001}.

We gratefully acknowledge the support of ANPCyT, Argentina; YerPhI, Armenia; ARC, Australia; BMWFW and FWF, Austria; ANAS, Azerbaijan; CNPq and FAPESP, Brazil; NSERC, NRC and CFI, Canada; CERN; ANID, Chile; CAS, MOST and NSFC, China; Minciencias, Colombia; MEYS CR, Czech Republic; DNRF and DNSRC, Denmark; IN2P3-CNRS and CEA-DRF/IRFU, France; SRNSFG, Georgia; BMBF, HGF and MPG, Germany; GSRI, Greece; RGC and Hong Kong SAR, China; ISF and Benoziyo Center, Israel; INFN, Italy; MEXT and JSPS, Japan; CNRST, Morocco; NWO, Netherlands; RCN, Norway; MNiSW, Poland; FCT, Portugal; MNE/IFA, Romania; MESTD, Serbia; MSSR, Slovakia; ARIS and MVZI, Slovenia; DSI/NRF, South Africa; MICIU/AEI, Spain; SRC and Wallenberg Foundation, Sweden; SERI, SNSF and Cantons of Bern and Geneva, Switzerland; NSTC, Taipei; TENMAK, T\"urkiye; STFC/UKRI, United Kingdom; DOE and NSF, United States of America.

Individual groups and members have received support from BCKDF, CANARIE, CRC and DRAC, Canada; PRIMUS 21/SCI/017, CERN-CZ and FORTE, Czech Republic; COST, ERC, ERDF, Horizon 2020, ICSC-NextGenerationEU and Marie Sk{\l}odowska-Curie Actions, European Union; Investissements d'Avenir Labex, Investissements d'Avenir Idex and ANR, France; DFG and AvH Foundation, Germany; Herakleitos, Thales and Aristeia programmes co-financed by EU-ESF and the Greek NSRF, Greece; BSF-NSF and MINERVA, Israel; Norwegian Financial Mechanism 2014-2021, Norway; NCN and NAWA, Poland; La Caixa Banking Foundation, CERCA Programme Generalitat de Catalunya and PROMETEO and GenT Programmes Generalitat Valenciana, Spain; G\"{o}ran Gustafssons Stiftelse, Sweden; The Royal Society and Leverhulme Trust, United Kingdom.

In addition, individual members wish to acknowledge support from CERN: European Organization for Nuclear Research (CERN PJAS); Chile: Agencia Nacional de Investigaci\'on y Desarrollo (FONDECYT 1190886, FONDECYT 1210400, FONDECYT 1230812, FONDECYT 1230987); China: Chinese Ministry of Science and Technology (MOST-2023YFA1605700), National Natural Science Foundation of China (NSFC - 12175119, NSFC 12275265, NSFC-12075060); Czech Republic: Czech Science Foundation (GACR - 24-11373S), Ministry of Education Youth and Sports (FORTE CZ.02.01.01/00/22\_008/0004632), PRIMUS Research Programme (PRIMUS/21/SCI/017); EU: H2020 European Research Council (ERC - 101002463); European Union: European Research Council (ERC - 948254, ERC 101089007), Horizon 2020 Framework Programme (MUCCA - CHIST-ERA-19-XAI-00), European Union, Future Artificial Intelligence Research (FAIR-NextGenerationEU PE00000013), Italian Center for High Performance Computing, Big Data and Quantum Computing (ICSC, NextGenerationEU); France: Agence Nationale de la Recherche (ANR-20-CE31-0013, ANR-21-CE31-0013, ANR-21-CE31-0022, ANR-22-EDIR-0002), Investissements d'Avenir Labex (ANR-11-LABX-0012); Germany: Baden-Württemberg Stiftung (BW Stiftung-Postdoc Eliteprogramme), Deutsche Forschungsgemeinschaft (DFG - 469666862, DFG - CR 312/5-2); Italy: Istituto Nazionale di Fisica Nucleare (ICSC, NextGenerationEU), Ministero dell'Università e della Ricerca (PRIN - 20223N7F8K - PNRR M4.C2.1.1); Japan: Japan Society for the Promotion of Science (JSPS KAKENHI JP21H05085, JSPS KAKENHI JP22H01227, JSPS KAKENHI JP22H04944, JSPS KAKENHI JP22KK0227); Netherlands: Netherlands Organisation for Scientific Research (NWO Veni 2020 - VI.Veni.202.179); Norway: Research Council of Norway (RCN-314472); Poland: Ministry of Science and Higher Education (IDUB AGH, POB8, D4 no 9722), Polish National Agency for Academic Exchange (PPN/PPO/2020/1/00002/U/00001), Polish National Science Centre (NCN 2021/42/E/ST2/00350, NCN OPUS nr 2022/47/B/ST2/03059, NCN UMO-2019/34/E/ST2/00393, UMO-2020/37/B/ST2/01043, UMO-2021/40/C/ST2/00187, UMO-2022/47/O/ST2/00148, UMO-2023/49/B/ST2/04085); Slovenia: Slovenian Research Agency (ARIS grant J1-3010); Spain: Generalitat Valenciana (Artemisa, FEDER, IDIFEDER/2018/048), Ministry of Science and Innovation (MCIN \& NextGenEU PCI2022-135018-2, MICIN \& FEDER PID2021-125273NB, RYC2019-028510-I, RYC2020-030254-I, RYC2021-031273-I, RYC2022-038164-I), PROMETEO and GenT Programmes Generalitat Valenciana (CIDEGENT/2019/023, CIDEGENT/2019/027); Sweden: Swedish Research Council (Swedish Research Council 2023-04654, VR 2018-00482, VR 2022-03845, VR 2022-04683, VR 2023-03403, VR grant 2021-03651), Knut and Alice Wallenberg Foundation (KAW 2018.0157, KAW 2018.0458, KAW 2019.0447, KAW 2022.0358); Switzerland: Swiss National Science Foundation (SNSF - PCEFP2\_194658); United Kingdom: Leverhulme Trust (Leverhulme Trust RPG-2020-004), Royal Society (NIF-R1-231091); United States of America: U.S. Department of Energy (ECA DE-AC02-76SF00515), Neubauer Family Foundation.

%
%


%

%
%
%

%
%
%
%

%
%
%

%
%
%
\clearpage
\printbibliography
\clearpage
\input{atlas_authlist}
%

%
%

%

%
%
%
%

\end{document}

%% file: atlas_authlist.tex
 
\begin{flushleft}
\hypersetup{urlcolor=black}
{\Large The ATLAS Collaboration}

\bigskip

\AtlasOrcid[0000-0002-6665-4934]{G.~Aad}$^\textrm{\scriptsize 103}$,
\AtlasOrcid[0000-0001-7616-1554]{E.~Aakvaag}$^\textrm{\scriptsize 16}$,
\AtlasOrcid[0000-0002-5888-2734]{B.~Abbott}$^\textrm{\scriptsize 121}$,
\AtlasOrcid[0000-0002-0287-5869]{S.~Abdelhameed}$^\textrm{\scriptsize 117a}$,
\AtlasOrcid[0000-0002-1002-1652]{K.~Abeling}$^\textrm{\scriptsize 55}$,
\AtlasOrcid[0000-0001-5763-2760]{N.J.~Abicht}$^\textrm{\scriptsize 49}$,
\AtlasOrcid[0000-0002-8496-9294]{S.H.~Abidi}$^\textrm{\scriptsize 29}$,
\AtlasOrcid[0009-0003-6578-220X]{M.~Aboelela}$^\textrm{\scriptsize 44}$,
\AtlasOrcid[0000-0002-9987-2292]{A.~Aboulhorma}$^\textrm{\scriptsize 35e}$,
\AtlasOrcid[0000-0001-5329-6640]{H.~Abramowicz}$^\textrm{\scriptsize 153}$
% ...
\bigskip
\\

% ...
$^{16}$Department for Physics and Technology, University of Bergen, Bergen; Norway.\\
% ...
$^{29}$Physics Department, Brookhaven National Laboratory, Upton NY; United States of America.\\
% ...
$^{35}$$^{(a)}$Facult\'e des Sciences Ain Chock, Universit\'e Hassan II de Casablanca;$^{(b)}$Facult\'{e} des Sciences, Universit\'{e} Ibn-Tofail, K\'{e}nitra;$^{(c)}$Facult\'e des Sciences Semlalia, Universit\'e Cadi Ayyad, LPHEA-Marrakech;$^{(d)}$LPMR, Facult\'e des Sciences, Universit\'e Mohamed Premier, Oujda;$^{(e)}$Facult\'e des sciences, Universit\'e Mohammed V, Rabat;$^{(f)}$Institute of Applied Physics, Mohammed VI Polytechnic University, Ben Guerir; Morocco.\\
% ...
$^{44}$Physics Department, Southern Methodist University, Dallas TX; United States of America.\\
% ...
$^{49}$Fakult\"{a}t Physik , Technische Universit{\"a}t Dortmund, Dortmund; Germany.\\
% ...
$^{55}$II. Physikalisches Institut, Georg-August-Universit\"{a}t G\"ottingen, G\"ottingen; Germany.\\
% ...
$^{103}$CPPM, Aix-Marseille Universit\'e, CNRS/IN2P3, Marseille; France.\\
% ...
$^{117}$$^{(a)}$New York University Abu Dhabi, Abu Dhabi;$^{(b)}$United Arab Emirates University, Al Ain; United Arab Emirates.\\
% ...
$^{121}$Homer L. Dodge Department of Physics and Astronomy, University of Oklahoma, Norman OK; United States of America.\\
% ...
$^{153}$Raymond and Beverly Sackler School of Physics and Astronomy, Tel Aviv University, Tel Aviv; Israel.\\
% ...
$^{a}$ Also Affiliated with an institute covered by a cooperation agreement with CERN.\\
$^{b}$ Also at An-Najah National University, Nablus; Palestine.\\
$^{c}$ Also at Borough of Manhattan Community College, City University of New York, New York NY; United States of America.\\
$^{d}$ Also at Center for Interdisciplinary Research and Innovation (CIRI-AUTH), Thessaloniki; Greece.\\
$^{e}$ Also at Centro Studi e Ricerche Enrico Fermi; Italy.\\
$^{f}$ Also at CERN, Geneva; Switzerland.\\
% ...

\end{flushleft}
